\documentclass[runningheads]{llncs}
\usepackage{graphicx}
\usepackage{float}
\usepackage[caption=false]{subfig} 
\usepackage{url}
\usepackage{todonotes}
\usepackage{xcolor}
\usepackage{float}
\usepackage{subfig}

\begin{document}
\mainmatter              
\title{Protecting Personal Data using Smart Contracts}
\titlerunning{}  

\author{Mohsin Ur Rahman$^{1}$, Fabrizio Baiardi$^{2}$, Barbara Guidi$^{1}$ and Laura Ricci$^{1}$ 
}

\authorrunning{Mohsin Ur Rahman et al.} 
%
\tocauthor{Mohsin Ur Rahman, Barbara Guidi, Laura Ricci, Fabrizio Baiardi}
\institute{Department of Computer Science, University of Pisa\\
\email{\{mohsinur.rahman,guidi,laura.ricci\}di.unipi.it}
\and
Department of Philology, Literature, and Linguistics \\ University of Pisa \\
\email{fabrizio.baiardi@unipi.it}}

\maketitle
\thispagestyle{empty}
\pagestyle{empty}

\begin{abstract}
Decentralized Online Social Networks (DOSNs) have been proposed as an alternative solution to the current centralized Online Social Networks (OSNs). Online Social Networks are based on centralized architecture (e.g., Facebook, Twitter, or Google+), while DOSNs do not have a service provider that acts as central authority and users have more control over their information. Several DOSNs have been proposed
during the last years. However, the decentralization of the OSN requires efficient solutions for protecting the privacy of users, and to evaluate the trust between users. Blockchain represents a disruptive technology which has been applied to several fields, among these also to Social Networks. In this paper, we propose a manageable, user-driven and auditable access control framework for DOSNs using blockchain technology. In the proposed approach, the blockchain is used as a support for the definition of privacy policies. The resource owner uses the public key of the subject to define flexible role-based access control policies, while the private key
associated with the subject's Ethereum account is used to decrypt the private data once access permission is validated on the blockchain. We evaluate our solution by exploiting the Rinkeby Ethereum testnet to deploy the smart contract, and to evaluate its performance. Experimental results show the feasibility of the proposed scheme in achieving auditable and user-driven access control via smart contract deployed on the Blockchain.

\end{abstract}


\section{Introduction}
In today's digital world, online users are generating a large amount of information. Online users are actively engaged on various sites including Facebook, Twitter, Instagram, LinkedIn, etc. In particular, a large amount of personal sensitive information are generated on popular social networking sites.
\sloppy Privacy is the main issue of current Social Network platforms, and several privacy disclosure events have happened, such as Cambridge Analytica\footnote{https://www.theguardian.com/technology/2019/mar/17/the-cambrid
ge
-analytica-scandal-changed-the-world-but-it-didnt-change-facebook}.
To face the privacy issue, decentralization has been proposed as a possible solution. A Decentralized Online Social Network (DOSN) is an Online Social Network (OSN) implemented in a distributed environment. During the last years, not only several academic solutions have been proposed \cite{guidi2015didusonet,LifeSocial10,EURECOM+2908}, but also online services such as Mastodon\footnote{https://mastodon.social/about} and Diaspora\footnote{https://diasporafoundation.org/}.

Thanks to the decentralization of data, users have more control over their data. However, decentralization introduces important requirements which have to be met, such as data availability and the information diffusion \cite{survey}.
As concerns privacy, access control techniques are used to manage the access to data in several fully-decentralized DOSNs \cite{bodriagov2014access}. Access control policies are used to express the rights of subjects to access services, and these policies are evaluated at access request time. For instance, in role-based access control system \cite{rbac}, policies are associated with resources in the network and users or groups of users are assigned roles, which determines their applicability to access the protected resources.
A DOSN requires that access control policies should be evaluated by users in the network to check if a user can access to a data or not \cite{de2017privacy}. Considering the dynamism of DOSNs due to the online/offline status of users \cite{comcom}, and the trust problem of choosing users which are not malicious, the validation of privacy policies is a tricky problem, which can be managed by using blockchain technology.

Blockchain technology emerged to the world due to Bitcoin, which was the first global cryptocurrency introduced in 2008 \cite{Barber}. It is worthy to note that the basic purpose of blockchain was to enable the functionality of a peer-to-peer (P2P) payment system without relying on a third party. In other words, Blockchain is a P2P network for conducting transactions in a secure and transparent manner. 

A blockchain is principally a list of records, called blocks.
Each block contains the hash of the previous block, and other essential information such as timestamp, and a set of transactions etc. Note that the first block of the chain is called a genesis block, which provides a foundation to create other blocks on top of it. 


Ethereum has received significant attention from industry and academia in recent years thank to the introduction of the smart contracts. Because of their 
resilience to tampering, smart contracts are used in many scenarios such as transfer of money, games etc. \cite{bartoletti2017empirical}.
Ethereum supports two types of accounts. Externally Owned Accounts (EOAs) are used to transfer money from one account holder to another one. Furthermore, each EOA is assigned a unique 20 bytes address, which uniquely identifies the account holder in the network.
Ethereum also supports contract accounts simply called contracts. Each contract account is associated with a unique code to uniquely identify it in the network.
Ethereum accounts are associated with unique public/private key pairs. Transactions can be sent to specific addresses represented by public keys, and only users with the specified destination addresses can access them in the network. 
The execution of transactions in the Ethereum network requires the payment of Ether. Thus, users possessing sufficient Ether can only execute the code of the smart contracts stored on the blockchain. Therefore, Ethereum users are required to purchase gas by paying Ether.
The consensus technique used by Ethereum is called Proof-of-Work, which enables miners to solve a cryptographic challenge (i.e., to guess random numbers). 


In this paper, we propose a blockchain-based access control for DOSNs. The role-based access control policies are stored on the blockchian, are publicly auditable and permit the verification of the user's rights even when the owner of the data is not logged in the social network. Our decentralized access control management system relies on the DOSN users to grant, revoke or update the access rights by making transactions to execute the respective functions of smart contract. DOSN users are associated with Ethereum accounts in order to uniquely identify themselves in the network. We focus on the Role-based Access Control Model (RBAC) \cite{rbac} because roles play a vital role in managing the contents of DOSNs. Thus, a DOSN user can send transactions to the access control contract deployed on the blockchain to assign roles to his/her colleagues, to family's and friends' members etc., and to allow only the intended users to access the resources of the resource owner. 

The rest of this paper is structured as follows: Section 2
is dedicated to the presentation of the background on DOSNs and of the Related Works. In section 3, we discuss the proposed framework. Section 4 presents the results of the performance evaluation. Finally, section 5 concludes the paper. 

\section{Background and Related Work}

Decentralized Online Social Networks (DOSNs) are introduced to overcome privacy issues of current OSNs. DOSNs offer a new revolution for data management, thereby allowing the users to control and manage their own personal information. In other words, DOSNs function as a dynamic peer-to-peer (P2P) network to store the users' profiles without reliance on a single service provider. However, DOSNs introduce many new problems regarding the availability, access control and availability of the shared items \cite{DOSN}. Various architectures for decentralized Online Social Networks have been proposed \cite{guidi2015didusonet,LifeSocial10,EURECOM+2908,buchegger2009peerson}. Usually encryption operations are used on the stored data such that only clients with the corresponding decryption keys are allowed to decrypt and view the stored contents. Access control is one of the main used technique to manage the access to data in DOSN. Indeed, DOSNs use privacy policies with encryption techniques, 
in many cases they use the Attribute Based Encryption (ABE) \cite{de2017privacy}. 




In detail, LifeSocial.KOM \cite{LifeSocial10}, and PeerSoN \cite{buchegger2009peerson}) are based on encryption. In particular, they combine different encryption schemes in order to protect the privacy of users' contents. Each content generated by a user should be encrypted and replicated on different users' devices. DOSNs, such as My3 \cite{narendula2011my3} and DiDuSoNet \cite{guidi2015didusonet}, do not rely on encryption techniques, indeed contents remains unencrypted on the devices, and trust is exploited to store data. 
The privacy solutions adopted by DOSNs must allow user to deny access to unauthorized contacts or to grant access to new contacts, regardless of whether their are based on encryption or not.

One of the main problem in DOSNs is that data are stored on users’ devices, so they are available as long as users are online. To enhance data availability, users' data are replicated on the devices of different users. It is also worth noticing that the evaluation of privacy policies regarding the data of a user can be evaluated by the user itself if she/he is online, otherwise it must be executed by some trusted node hosting the data replica.

Blockchain technology is a disruptive solution which can be exploited by DOSNs in order to improve privacy systems.
During the last years, blockchain has been applied to several scenarios and also to Social Networks. Several social Network platforms \cite{steem,Synereo,sapien} exploiting blockchain have been proposed mainly in order to overcome the privacy problems of current OSNs. In many cases the blockchain is exploited to support techniques for the detection of  fake news. 

However, to the best of our knowledge, all current Blockchain-based Social Networks do not exploit blockchain as a support to control accesses to the data owned by the users.

\section{The Proposed Framework}
The basic objective of our proposal is to use the blockchain to store and evaluate role-based access control policies. To achieve this objective, the resource owner can send transactions to the role-based access control smart contract deployed on the Ethereum blockchain in order to assign/evaluate roles to other DOSN users in the network.
The access control policies are stored on the blockchain, thereby allowing the resource owner as well as other nodes to check the enforcement of the privacy policies when the contract receives access requests. 
Our framework is privacy preserving because the real identifies of DOSN users are not disclosed to other users in the network. The main rationale behind our approach is motivated from the fact that Ethereum uses a single public address for each account holder. 
We need to make the following essential configurations to apply the Ethereum platform in our access control system.  
\begin{itemize}
\item Each DOSN user must be associated with an Ethereum account to uniquely identity itself in the network. Thus, this account allows each peer to claim the deployment of a smart contract, and to assign roles to other DOSN users
\item All DOSN users can configure and run the Ethereum client because such devices have sufficient computing capabilities. A DOSN user can use its Ethereum client to directly interact with the blockchain. Indeed, they can also send transactions to execute the main functions of smart contracts.
\end{itemize}

In our framework, a DOSN user can easily check the blockchain at any time to determine whether he/she is assigned a role from another DOSN user. Furthermore, this approach is intended to achieve distributed auditability, thereby preventing third parties from fraudulently denying the access rights granted by role-based access control policies. It is worthy to note that a  subject who wishes to access the resources of the resource owner must assert that it possesses a role that was issued by the resource owner. 
The architecture of our access control framework is depicted in Figure 1. The main actors of our framework include the resource owner, a node trusted from the resource owner and subjects (i.e., users who are interested in viewing the resources of the resource owner). The resource owner uses the Ethereum address of the subject in order to define role-based privacy policies using smart contract. 

\begin{figure*}[h]
\centering
\includegraphics[ width=0.7\columnwidth]{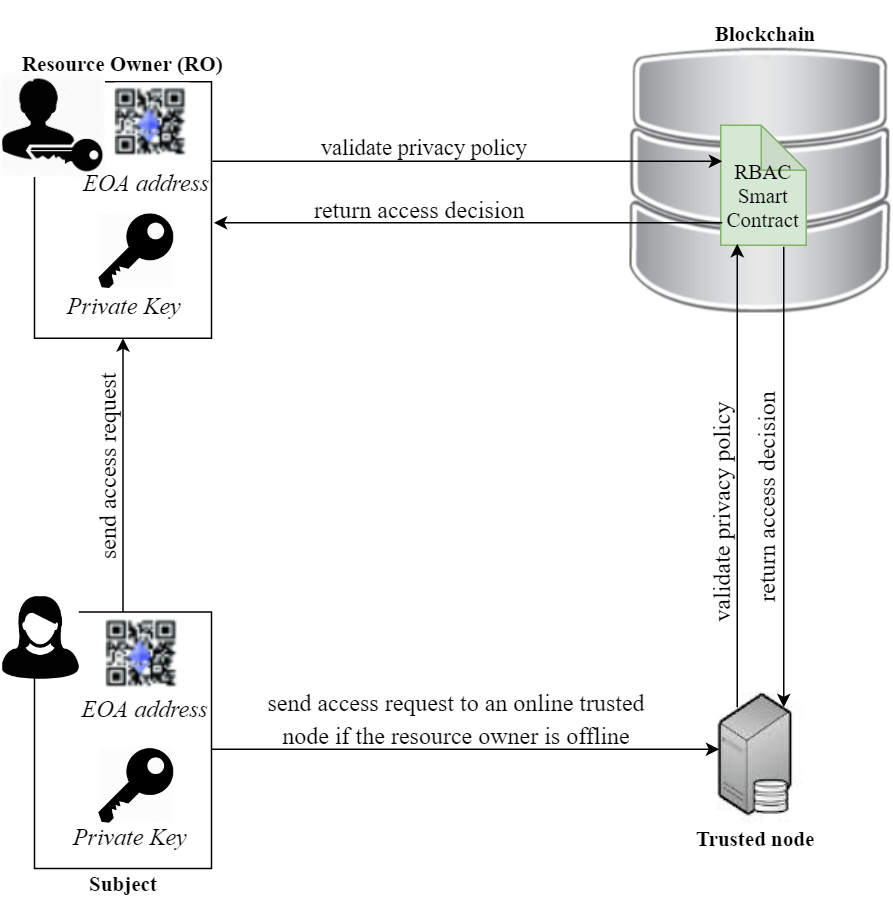}
\caption{Architecture of the proposed framework for DOSNs using Smart Contract}
\end{figure*}

In our approach, the resource owner has a set of trusted nodes where data are stored, as described in \cite{de2017privacy}, in order to guarantee the data availability even when the user is offline. The choice of trusted replica nodes is outside the scope of this article. 
A generic request is sent to the resource owner when it is online, otherwise it is sent to one trusted node chosen at random. These nodes check the privacy policies to determine the permission associated with roles. Based on the evaluation of access control policies, the subject is either allowed or denied to view or download the private data of the resource owner. 

When a new relationship is established between the resource owner and the subject, the later involved in the relationship has to share its public key (i.e., Ethereum address). After verification of the user's identity, the resource owner sends a transaction to the access control contract in order to issue a desired role such as colleague, friend, friend-of-friend or family etc. 
Finally, after the establishment of a relationship and privacy policy definition using blockchain, the subject can send access requests to the resource owner or to trusted nodes when the former is offline. 
Finally, after confirming information in the blockchain, the resource owner or trusted node is assured that the asserted role is assigned by the resource owner. 






\subsection{Main Functions of the Framework}
The resource owner can send transactions to invoke the main functions of the RBAC deployed on the blockchain. 
An overview of the main functions of the proposed framework is given below;
\begin{itemize}
    \item {\bf{Policy Creation}:}
   This function allows the resource owner to issue role to a requesting DOSN user in the network. It requires the Ethereum address of the subject, and the role that the resource owner wishes to assign. As a simple example, if Bob (i.e., the resource owner) wishes to add Alice (i.e., subject) as friend, he simply needs to provide the Ethereum address of Alice, and the role friend and an access permission. 
   To accomplish this goal, the resource owner simply needs to make a new policy creation transaction to execute the \emph{policyAdd()} function of the RBAC smart contract deployed on the blockchain. 
    \item {\bf{Policy Update}}:
  In the proposed framework, the resource owner can update a policy at anytime. The resource owner needs to specify the new role, permission and address of the subject for which the privacy policy needs to be updated. To accomplish this goal, the resource owner simply needs to make a policy update transaction to execute the \emph{policyUpdate()} function of the RBAC smart contract. This function requires address of the subject, role and permission as parameters to assign a new role or change the associated access permission.  
  \item {\bf{Policy Deletion}}: The resource owner can also revoke an existing privacy policy by making a policy revocation transaction to execute the \emph{policyDelete()} function of the RBAC smart contract. 
  \item {\bf {Right Transfer}}:
  The resource owner must not only be able to specify who can access the resource, he must also be able to add users who can further delegate that authorization to other users. To accomplish this goal, only users endorsed by the resource owner can send transactions to execute the \emph{roleTransfer()} function of the smart contract. This function requires the link or identification information of the policy and the address of the new user to transfer the access rights. The framework is scalable because the current right holders can further transfer the access rights to new users as well. 
  
  \item {\bf{Access Control}}: This function allows the resource owner or trusted nodes to check access permission on the blockchain when an access request is submitted by the subject. To accomplish this goal, the resource owner or trusted nodes perform an access control transaction to execute the \emph{AccessControl()} function of the smart contract. 
  
\item {\bf{Contract Deletion}}: The resource owner can also disable the RBAC smart contract by making a contract deactivation transaction to execute the \emph{deleteRBAC()} function of the RBAC smart contract. 
\end{itemize}

We use two-dimensional mappings from the elements of address (primary key) and role (secondary key) to create, update, revoke and validate these policies in the blockchain.

\subsection{Challenge-Response Authentication}
The challenge-response protocol is executed when a subject sends an access request to the resource owner or to a trusted node when the former is offline. The basic steps of this protocol are discussed as follows;
\begin{itemize}
    \item \textbf{Declaration:} The aim of this step is to enable
the subject to declare that it possesses a
unique Ethereum account/address that was used to issue a role, and permission by
the resource owner. In our simple example, Alice will assert
the role friend to indicate that this role has been issued by
Bob in order to access his resources. Thus, the asserted role makes Alice eligible to access the
protected resources of Bob.

\item \textbf{Information Verification:} Given the assertion of the subject, the resource owner or trusted node checks the RBAC policies published on the
blockchain to determine weather the corresponding public key was issued a
role by the resource owner. Given the data on the blockchain, these nodes
will be able to check the information related to the public key of the subject.
Finally, the node receiving the request challenges the requesting DOSN user
in order to verify his/her real identity (i.e., to determine whether he/she is
the true owner of the Ethereum account).
   
\item \textbf{Challenge:} The resource owner or trusted node selects a random data $d$, and asks the subject to sign it.

\item \textbf{Response:} Response from the subject determines the true owner of the corresponding Ethereum account
because the private key is required to sign the message.
Therefore, the creation of a correct signature is only possible
if the user possesses the corresponding private key. Thus, the message can only be signed by the legitimate owner. To accomplish this goal, we use the following function:
\begin{equation} 
S = Sign(pk; address; d)
\end{equation}

where $pk$
represents the private key of the subject, and
$d$ represents the random data sent to sign. Thus, a correct
signature is only possible if the subject possesses the
corresponding private key which is uniquely defined for each
Ethereum account, and it is also unique for each DOSN user. The subject then sends $S$ back to the resource owner or trusted node.

\item \textbf{Response Confirmation:} The resource owner or trusted nodes will allow
the subject to access the protected resources if
and only if a correct signature is generated by the subject. To accomplish this goal, the resource owner or trusted node uses the following verification function:
\begin{equation}
    confirmResponse(address; d; S).
\end{equation}
\end{itemize}

Finally, the resource owner or trusted node will allow access to the resource if and only if the verification process is successful. Please note that the authentication protocol can be executed offline, and the Solidity sha3 \cite{sha3}  function can be used to securely generate the signature. This technique can be used to generate a message signature without disclosing the private key. 
\section{Performance Evaluation}
Every transaction that is used to invoke the function of a smart contract requires the payment of a fee in order to compensate the mining node for the execution of transaction and saving it on the blockchain. Ethereum uses gas to express this fee. Users can purchase gas from the mining nodes by paying Ether. Please note that Gas and Ether are two distinct terms because gas indicates a constant cost of performing an operation on a Blockchain network, whereas Ether is a volatile virtual currency, which is used to pay for the network resources.
\subsection{Experiment Setting}
We use the solidity \cite{sc} programming language to develop a prototype of the proposed RBAC smart contract, and deployed it on the Rinkeby Ethereum testnet. During analysis in the month of April, we observed an average gas value of $\approx$ 0.000000021 ETH, and 1 Ether $\approx$ 137.66 USD. Experimental results show that our proposed smart contract requires 1869303 gas for deployment. Therefore, the creation and deployment of our proposed smart contract on the blockchain requires 0.256 USD. However, it is only a single time cost to initialize our proposed RBAC smart contract. 

\subsection{Results}
Table 1 shows the one-time costs of the functions of the proposed smart contract when a subject is assigned a role by the resource owner. It is worthy to note that that a role is assigned to the user together with an additional permission element. As it can be observed from the table, the gas consumption costs are slightly increased. However, the costs of the remaining three functions are always constant because these functions are independent of the main functions of the smart contract (i.e., policy creation and policy updation). 

Table 1 also shows that the one-time constant cost of the access control function is 0.0031 USD\$. The resource owner can also delete the smart contract from the blockchain by invoking the \emph{deleteRBAC()} function, which performs the suicide or self-destruct operation to remove the code of RBAC from the Ethereum blockchain \cite{suicide}. 

\begin{table}
\centering
\caption{Costs of the different functions of the RBAC smart contract }\label{tab1} 
\begin{tabular}{|l|l|l||l|}
\hline
RBAC Function &  Gas Used & Cost(ether) & USD(\$)\\
\hline
\emph{policyAdd()} &  27864 & 0.000028 & 0.0038\\
\emph{policyUpdate()} &  27800 &  0.000028 & 0.0038‬\\
\emph{policyDelete()} &  22680 & 0.000023 & 0.0031\\
\emph{roleTransfer()} &  51456 & 0.000051 & 0.0069\\
\emph{AccessControl()}  & 22808 & 0.000023 & 0.0031\\
\emph{deleteRBAC()}  &  13455 &  0.000027 &  0.0036\\
\hline
\end{tabular}
\end{table}
We performed an experiment to evaluate the effects of the number of subjects in the RBAC policy on Gas consumption (i.e., when the resource owner assigns role to a group of subjects in a single policy creation transaction). The results of the experiment are presented in Figure 2(a). Results show that the policy creation and update transaction for 1 Ethereum address require 3,392 gas consumption on the Rinkeby testnet and this cost increases linearly as we increase the number of addresses on these two transactions. We also observed that the contract deployment gas cost also increases linearly as the number of users in the RBAC policy are increased. It is worthy to note that the given results show the applicability of the proposed framework to simultaneously assign the same role to multiple users in a single policy creation transaction.    

We performed another experiment to evaluate the effects of the number of bits on gas consumption as shown in Figure 2(b). Results of the experiment show that each single bit denoting a particular input in the policy creation and updation transactions requires constant amount of gas on the Rinkeby testnet.
\raggedbottom
\begin{figure}%
    \centering
    \subfloat[Number of users vs Gas cost]{{\includegraphics[width=0.44\textwidth]{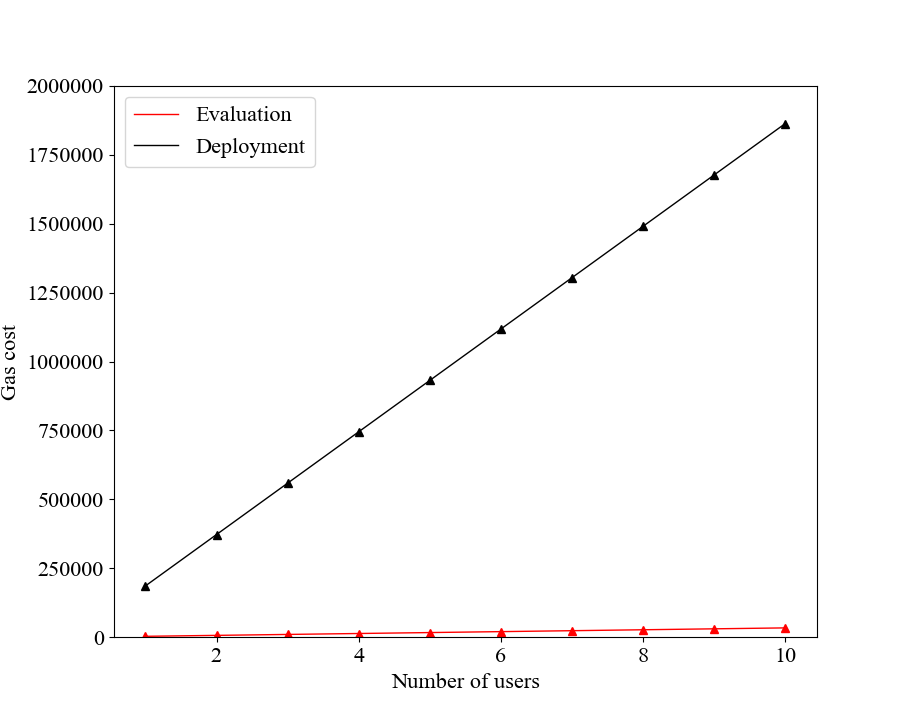} }}%
    \qquad
    \subfloat[Number of bits vs Gas cost]{{\includegraphics[width=0.46\textwidth]{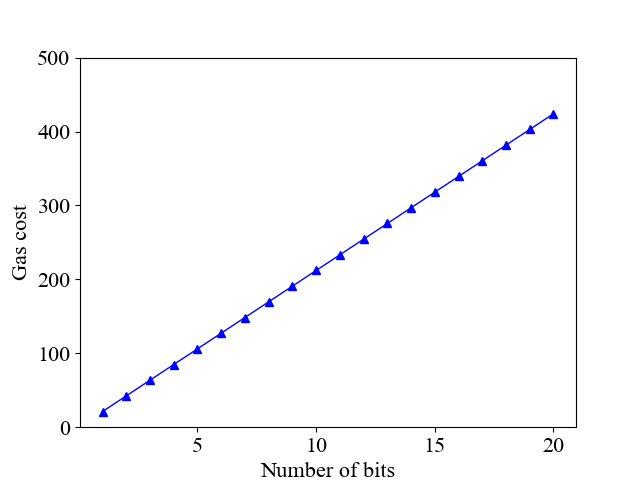} }}%
    \caption{Effects of number of users (a) and bits (b) on gas consumption}%
    \label{fig:example}%
\end{figure}
\raggedbottom

The privacy policies can be valid if miners have approved all the related transactions to the smart contract deployed on the blockchain, and these transactions are recorded to new blocks. It is worthy to note that the block generation time is directly proportional to the transaction rate (i.e., the more transactions are made, the more time it will take to generate the relevant block). We setup our own private Ethereum network to assess the effects of the number of miners on block generation time. Initially, we configure one miner to mine block. However, as we increased the number of miners, the block generation time decreased and became stable when the number of miners reached to 4 as shown in Figure 3. Furthermore, we also confirmed the block generation time on Rinkeby testnet.  
\begin{figure}[H]
\centering
\includegraphics[width=0.6\columnwidth]{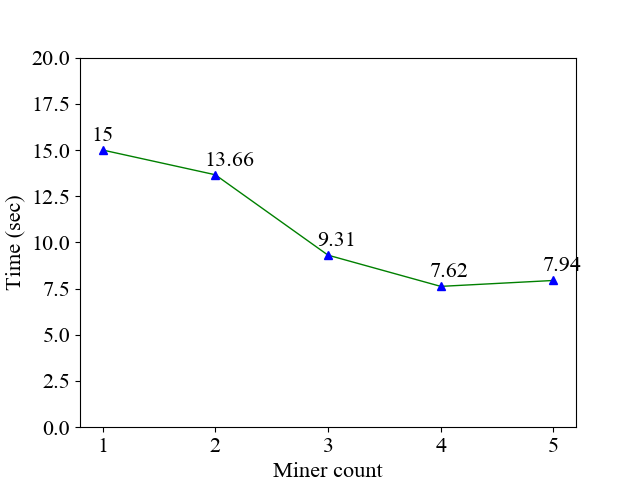}
\caption{Block Generation Time vs. Number of Miners}
\end{figure}
\subsection{Security Analysis}
\raggedbottom
It is not possible for an adversary to issue role-based policies on behalf of the resource owner because execution of the contract functions is only allowed to the owner of the RBAC smart contract. Furthermore, acting as
a subject, the goal of the attacker is to impersonate a
role that it does not possess (i.e., it can try to assert false roles to the resource provider or trusted nodes) in order to gain illegal access to the resources of another DOSN user. However, as already discussed, each DOSN user has unique private key in the network, and an adversary is unable to compromise the private keys of the users in the system.


Each user in the DOSN network can chose one or more of its neighbors as trusted nodes. These nodes are selected to perform trusted actions in the network including data storage and access control. 

We assume that the cryptographic primitives used in the
framework can not be broken by the adversary. Therefore, they
are unable to forge the digital signature and to obtain hash
collisions. It is worthy to note that the proposed framework is
blockchain-oriented, thus, we assume that an adversary can not control a majority of the computing power in the network in order to avoid the 51\% attack.

The challenge-response protocol is used to perform a secure verification of the user's identity, role and the associated access permission on the blockchain. The verification steps used in the challenge response protocol can be conducted offline, and latest technologies used in devices such as Near Field Communication (NFC) and Quick Response (QR) codes \cite{want2011near} can be used to effectively transfer the message and signature.


\begin{figure}[H]
\centering
\includegraphics[width=0.6\columnwidth]{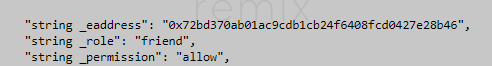}
\caption{The output of policy creation transaction using Remix IDE \cite{sc}} \raggedbottom
\end{figure}
\begin{figure}[H]
\centering
\includegraphics[width=0.6\columnwidth]{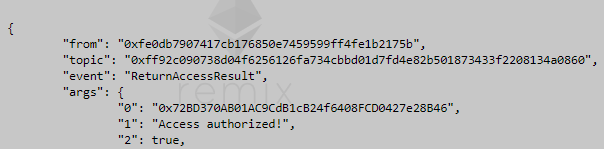}
\caption{The output of privacy policy validation transaction}
\end{figure}
\section{Conclusion and Future Work}
In this research, we proposed an auditable and user-driven role-based access control management framework (RBAC) for DOSNs using blockchain technology.
DOSN users are associated with Ethereum accounts having unique public, private keys. The public keys serve as addresses of the users while the private keys
are used to prove the real identities of DOSN users. Each DOSN user is capable
to invoke the assign and revoke functions of the access control contract deployed
on the blockchain. Advantages of the proposed scheme include auditability (i.e.,
the user-role assignment is visible on the blockchain). Thus, any user can easily
inspect them at any time. Furthermore, they can include other elements such as
expiry time to automatically revoke the role assigned to users. We are planning
to extend our work in several directions. We will investigate the possibility of
exploiting external storages, such as private clouds, to guarantee data availability for offline users. Furthermore, we are developing a DAPP to interact with a
system of different smart contracts on the blockchain including a tool to translate role based access control policies to smart contracts to be deployed on the
blockchain.

%
%
%
\bibliographystyle{splncs04}
\bibliography{mybibliography}

\end{document}